\documentclass[graybox]{svmult}

\usepackage{type1cm}         

\usepackage{makeidx}         
\usepackage{graphicx}        
\usepackage{multicol}        
\usepackage[bottom]{footmisc}

\usepackage{aa}              
\usepackage{chemformula}     

\usepackage{newtxtext}       %
\usepackage[varvw]{newtxmath}

\makeindex             

\begin{document}

\title*{Observational and computational characterization of exoplanet atmospheres}


\author{
Daniel Kitzmann\orcidID{0000-0003-4269-3311} \\
Elspeth K. H. Lee\orcidID{0000-0002-3052-7116} \\
Jens Hoeijmakers\orcidID{0000-0001-8981-6759} and \\
Kevin Heng\orcidID{0000-0003-1907-5910}
}

\authorrunning{Kitzmann, Lee, Hoeijmakers \& Heng}

\institute{
Daniel Kitzmann \at Space Research and Planetary Science, University of Bern, Gesellschaftsstrasse 6, 3012 Bern, Switzerland \email{daniel.kitzmann@unibe.ch}
\and Elspeth K. H. Lee \at Center for Space and Habitability, University of Bern, Gesellschaftsstrasse 6, 3012 Bern, Switzerland \email{elspeth.lee@unibe.ch}
\and Jens Hoeijmakers \at Lund Observatory, Division of Astrophysics, Department of Physics, Lund University, Box 118, 221 00 Lund, Sweden \email{jens.hoeijmakers@fysik.lu.se}
\and Kevin Heng \at Ludwig Maximilian University, University Observatory Munich, Scheinerstraße 1, 81679 München, Germany, \email{kevin.heng@physik.lmu.de}
}

%
%
\maketitle


\abstract*{This chapter reviews the current state of observational and theoretical efforts in the characterization of exoplanet atmospheres, with a focus on developments enabled through the Swiss National Centre for Competence in Research (NCCR) PlanetS. It covers the essential physical and chemical processes that govern atmospheric dynamics, radiative transfer, chemistry, and cloud formation in exoplanets and brown dwarfs. The review discusses the modeling approaches used to simulate these processes, ranging from simplified 1D models to fully coupled 3D general circulation models. Atmospheric retrieval frameworks are presented as tools for inferring atmospheric properties from observational data, highlighting both classical Bayesian techniques and emerging machine learning methods. Observational strategies using instruments like HST, JWST, and ground-based high-resolution spectrographs are also examined. Special emphasis is placed on the interplay between theory and observation, and how developments in modeling, data analysis, and instrumentation collectively advance our understanding of planetary atmospheres beyond the Solar System.}


\section{State of the field}
\label{sec:1}

The field of exoplanet science has rapidly evolved from the detection of exoplanets to the detailed characterization of their atmospheres. Thanks to significant advances in observational capabilities, we are now able to study worlds beyond our Solar System in unprecedented detail. Ground-based search campaigns such as WASP (Wide Angle Search for Planets) and NGTS (Next-Generation Transit Survey), alongside space-based missions like Kepler, TESS (Transiting Exoplanet Survey Satellite), and the upcoming PLATO (PLAnetary Transits and Oscillations of Stars), have uncovered a remarkably diverse population of exoplanets.

These observations have revealed a broad spectrum of planetary types, ranging from gas giants of varying sizes to rocky super-Earths and smaller terrestrial worlds. Many of these planets have no direct analogues in our own Solar System, making their characteristics and formation processes a major focus of ongoing research. Instruments such as the CHaracterising ExOPlanet Satellite (CHEOPS), the James Webb Space Telescope (JWST), and the Hubble Space Telescope (HST), together with large ground-based observatories like the Very Large Telescope (VLT) and the forthcoming Extremely Large Telescope (ELT), are instrumental in advancing this work.

Exoplanet spectra can generally be obtained through two primary methods. The first is during a transit, when the planet passes in front of its host star and its atmosphere can be studied in \textit{transmission} \cite{Seager2000ApJ...537..916S,Charbonneau2002ApJ...568..377C}. The second occurs around the time of the secondary eclipse, when the planet's day-side is visible, allowing its \textit{emission spectrum} to be isolated \cite{Deming2005Natur.434..740D}. Spectroscopic observations conducted over the entire orbit can produce a \textit{phase-curve} \cite{Stevenson2014Sci...346..838S}, i.e. the different phases of the irradiated planet orbiting its host star. If the planet's night-side is sufficiently hot and luminous, can also reveal the night-side spectrum \cite{Evans2022NatAs...6..471M}. Besides emitted thermal radiation, the emission spectrum may also contain contributions by light from the host star that is scattered by the planet.

Understanding how exoplanets form and evolve requires knowledge of their chemical compositions. One important diagnostic is the carbon-to-oxygen (C/O) abundance ratio, which is influenced by the location in the protoplanetary disk where a planet forms \cite{Oberg2011ApJ...743L..16O, Madhusudhan2012ApJ...758...36M}. However, accurately measuring the C/O ratio is challenging, and recent research has shown that a broader range of atmospheric constituents must be considered to better constrain a planet’s formation history \cite{Lothringer2021ApJ...914...12L}.

A significant proportion of the known exoplanets are gas giants orbiting very close to their host stars, reaching temperatures of up to $\sim$2000\,K, conditions not found in our Solar System. Aside from the extremely hot gas giants known as ultra-hot Jupiters, these planets exhibit complex molecular and condensation chemistry. At high temperatures, refractory elements form a variety of gas-phase molecules and condensates, which manifest as clouds or dust. These condensates remain a longstanding challenge for atmospheric studies from both a theoretical as well as observational perspective. They, for example, obscure our view of the deeper atmosphere and are difficult to detect and characterize observationally. Even with the unprecedented capabilities of JWST, clouds continue to pose a significant barrier to fully understanding exoplanet atmospheres.

Brown dwarfs share many physical properties with giant exoplanets, including similar radii and temperature ranges \cite{Marley2015ARA&A..53..279M}. As a result, they are expected to exhibit comparable atmospheric processes, particularly in terms of aerosol formation, including clouds and hazes \cite{Helling2008MNRAS.391.1854H,Marley2013cctp.book..367M}. Since the first detections of directly imaged exoplanets \cite{2004A&A_Chauvin, 2008Sci_Kalas}, studies have noted strong spectral and atmospheric similarities between these planets and brown dwarfs \cite{1995Natur_Rebolo, 1995_Oppenheimer}. Despite differences in mass, dynamical regimes of their atmosphere, or heating mechanisms, both exhibit similar spectral features across a wide range of temperatures, suggesting a shared evolutionary pathway in terms of dominant atmospheric chemistry, cloud structure, and radiative processes.

This similarity makes brown dwarfs valuable analogues for testing atmospheric models, as they can be studied in much greater detail. Unlike exoplanets, brown dwarfs are not affected by the brightness of a host star, resulting in cleaner, higher-quality data. This allows for more precise investigations into the physical and chemical mechanisms shaping their atmospheres. By studying brown dwarfs, we gain critical insights that enhance our understanding of the atmospheres of warm and cool exoplanets, including sub-Neptunes and gas giants.

\section{Modelling of exoplanet and brown dwarf atmospheres}

\begin{figure}[t]
\centering
\includegraphics[width=0.8\textwidth]{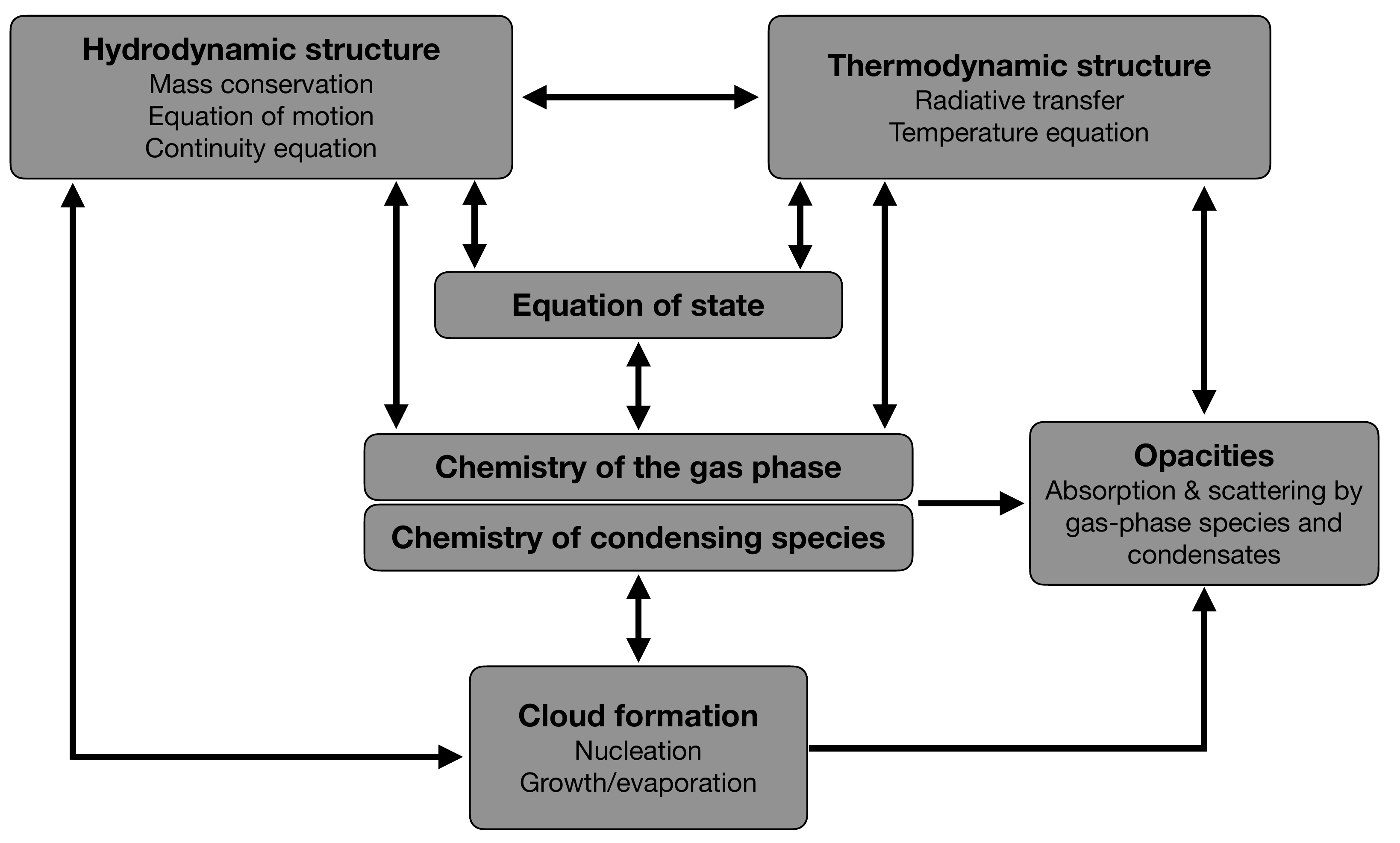}
\caption{Physical and chemical processes of self-consistent theoretical models of (exo)planetary and brown dwarf atmospheres. The double arrows indicate the mutual influence between the corresponding elements, reflecting the interplay of physical and chemical mechanisms considered within each component.}
\label{fig:atmosphere_flowchart}
\end{figure}

The complex challenge of modeling an exoplanet or brown dwarf atmosphere is illustrated in Figure \ref{fig:atmosphere_flowchart}. It involves a coupled system of equations that describe, for example, hydrodynamics, thermodynamics, chemistry, or cloud formation. When combined with appropriate boundary or initial conditions for the corresponding set of differential equations, the resulting mathematical framework forms a complete and well-posed system. Solving this system yields a \textit{theoretical model} of the atmospheric state, including key physical properties such as temperature and chemical composition, as well as its spectral characteristics.

A crucial requirement for a model to produce reliable predictions is \textit{self-consistency}. A self-consistent model ensures that all components represented in Figure \ref{fig:atmosphere_flowchart} are determined based on fundamental physical theories, and that the model outcomes result from solving the basic equations of the system rather than relying on arbitrary or ad hoc assumptions. For instance, in a self-consistent approach, cloud properties should emerge naturally from the model calculations through the solution of the relevant equations, rather than being predefined, for example, by assuming a fixed cloud particle size distribution.

However, self-consistency does not necessarily mean that every equation must be solved in its most general form. Models can be constructed with varying degrees of complexity, depending on the goals of the study and computational constraints. For example, a simplified model might assume a one-dimensional, hydrostatic atmosphere, whereas a more advanced model may incorporate full three-dimensional atmospheric dynamics.

In the following sections, we will briefly outline the key components required to construct a self-consistent model of an exoplanet atmosphere. A more detailed overview on the modeling of habitable, rocky planets is provided in Chapter 14.

\subsection{Opacities \& Radiative Transfer}

\subsubsection{Radiative transfer}
\label{sec:radiative_transfer}

Besides convection and atmospheric dynamics, radiative transfer is one of the most important energy transport mechanisms that determines the local energy budget, and thus the temperature, of any atmosphere. Furthermore, radiative transfer calculations provide the theoretical spectral appearance of an atmosphere and, thus, can connect the theoretical model with observations. Radiative transfer is consequently considered to be one of the fundamental ingredients in calculating planetary atmospheres but also in characterising observational data. 

For describing (exo)planet atmospheres, one usually assumes a plane-parallel, stationary atmosphere, which simplifies the radiative transfer to
\begin{equation}
    \mu \frac{\mathrm d I(\mu, z)}{\mathrm d z} = - \left(\kappa + s \right) I(\mu, z) + \kappa B + \frac{s}{2} \int_{-1}^{+1} p (\mu, \mu') I(\mu', z) \, \mathrm d \mu' \ .
    \label{eq:rt}
\end{equation}
Mathematically, this is a first-order integro-differential equation for the intensity $I$ as a function of the cosine of the polar angle $\mu$, the vertical distance $z$, the absorption ($\kappa$) and scattering ($s$) coefficients, the thermal emission with the Planck function $B$, and the scattering phase function $p$. Depending on the local physical conditions, this equation can have properties of an elliptic, a hyperbolic, and a parabolic differential equation, which together with the presence of the scattering integral, makes this equation difficult to solve in general \cite{Kanschat2009nmmr.book.....K}. 

\subparagraph{Improved two-stream methods}

However, in many cases one is not directly interested in the full intensity field $I$ but rather its angular moments, such as the mean intensity $J$ or the radiation flux $F$. It is only these average quantities that enter, for example, the temperature equations in an atmospheric model or that are used in observations, such as the radiation flux. Especially in the modelling of planetary atmosphere, it is, therefore, customary to not solve the full radiative transfer equation but rather its angular-averaged counterparts to directly calculate either the mean intensity or the flux.

One general framework of simplifying the transfer equation are the so-called two-stream methods. Instead of solving Equation \eqref{eq:rt} for all directions $\mu$, the transfer equation is re-written into a set of equations for the upward and downward flux or mean intensity. This greatly simplifies the solution and especially the required computational effort. However, this simplification comes at a price. The construction of a two-stream method usually leads to an open system of equations, meaning it always contains one more variable than there are equations. Thus, this system has to be artificially closed by making assumptions on, for example, the ratios of $J$ and $F$. Furthermore, the choice on how to simplify the scattering phase function in a two stream approach is also not unique. Two-stream methods, therefore, constitute a whole class of radiative transfer methods, each with their own assumptions on how to treat the angular scattering and on how to close the system of equations.

Despite their simplifications, these methods are pre-dominantly used in the modelling of planetary atmospheres. They are, especially, extensively employed in calculating the radiation field within a general circulation model, where solving the full Equation \eqref{eq:rt} is computationally unfeasible.

While the classic two-stream radiative transfer schemes (e.g. \cite{Meador1980JAtS...37..630M}, \cite{Toon1989JGR....9416287T}) often give reasonable results, they can also yield inaccurate solutions for specific scenarios. One such case is the back-scattering of infrared light by large dry ice cloud particles that were suggested by \cite{Forget1997Sci...278.1273F} to be able the provide a substantial greenhouse effect in the early Martian atmosphere. As demonstrated in \cite{Kitzmann2016ApJ...817L..18K} by using a multi-stream radiative transfer scheme (\cite{Stamnes1988ApOpt..27.2502S}), the study, based two-stream radiative transfer scheme, overestimated the surface warming due to the scattering greenhouse effect by up to 80\,K.

Therefore, \cite{Heng2017ApJS..232...20H} introduced a new generalisation of the two-stream radiative transfer description that can provide more accurate solutions of the radiative transfer equation in the presence of strong infrared scattering by large aerosols. In \cite{Heng2018ApJS..237...29H} this improved two-stream was further developed to also include a stellar beam source. The generalisation of the original two-stream equations employ a correction factor that is derived from more complex multi-stream radiative transfer solutions, allowing to treat the scattering of light in a more accurate way. At the same time, the improved two-stream methods still retain the simplicity of implementation of their classic counterparts, making them suitable to be used in three-dimensional general circulation models.

Traditional longwave 1D multiple-scattering radiative-transfer methods such as those mentioned in the previous paragraphs typically involve expensive matrix inversion operations as part of their methodology. 
Where computational efficiency is paramount, such as in retrieval and 3D GCM simulations, a trade-off between accuracy and speed must be made to remain viable in a suitable timeframe.
To tackle this, in \cite{Lee2024ApJ...965..115L} examined the approximate longwave radiative-transfer methods; Absorption Approximation (AA) (\cite{Li2000JAtS...57.2905L}) and Variational Iteration Method (VIM) (\cite{Zhang2017JAtS...74..419Z}) which were first developed for Earth atmospheric science, but now altered for a sub-stellar atmosphere context. \cite{Lee2024ApJ...965..115L} found that the AA and VIM methods were highly suitable and accurate for sub-stellar atmosphere calculations, producing similar heating/cooling rates in the atmosphere and outgoing fluxes to traditional inversion methods but with a significant speed up in computational time.

\subparagraph{Cloudy Monte Carlo Radiative Transfer}

While the two-stream methods discussed above are extensively used to compute temperature structures within general circulation models, these simplified radiative transfer schemes are less well suited to provide the spectral appearance in terms of reflected and emitted light of the three-dimensional atmosphere. However, in the current era of large ground-based telescopes and the capabilities of the James Webb Space Telescope the observational data of exoplanets increasingly provide detailed information on their 3D atmospheric structures. In order to compare the GCM simulations with actual observations, it is thus necessary to accurately calculate theoretical emission, albedo, and transmission spectra from the output of general circulation models. 

In order to account for all the light that is emitted and scattered towards the observer, one in theory has to solve the full three-dimensional version of the radiative transfer equation. Even doing so just as a postprocess procedure is computationally very demanding. A much more efficient way is available through the use of a Monte Carlo radiative transfer that can also deal with non-isotropic scattering by cloud particles. Such a Cloudy Monte Carlo Radiative Transfer (CMCRT) was introduced by \cite{Lee2019MNRAS.487.2082L}. This radiative transfer code is able use GCM output and transform it into the corresponding emission and transmission spectra. It was tested against more complex radiative transfer models and found to working be working excellent in most cases, while being much faster. The Monte Carlo radiative transfer code is, therefore, highly suitable for detailed post-processing of cloudy and non-cloudy 1D and 3D exoplanet atmosphere simulations in instances where atmospheric inhomogeneities or significant limb effects are important to explain observational data.

In \cite{Lee2022ApJ...929..180L} the model was further improved to run on GPUs rather than CPUs, which strongly increased the computational performance of the code. The new gCMCRT model has been specifically designed to be used directly on output of many popular general circulation model, including THOR, the UK Met Office UM, or Exo-FMS and has been released as open-source software\footnote{\url{https://github.com/ELeeAstro/gCMCRT}}. It also includes the option to postprocess the spectra at high spectral resolution, which makes it applicable in the interpretation of exoplanet data obtained with high-resolution spectrographs on large ground-based telescopes, such as CRIRES+ on the Very Large Telescope (VLT), for example.

\subsubsection{Opacities}

Opacities describe the interaction of light with matter and are one of the fundamental inputs for atmospheric models. Although the term \textit{opacity} is not uniquely defined across different fields, it generally encompasses both absorption and scattering coefficients.

The shape of an absorption line is typically a convolution of two effects: thermal (Doppler) broadening, which produces a Gaussian profile, and pressure broadening, described by a Lorentzian profile. The resulting line shape is mathematically described by a Voigt profile. To accurately model these features, both the line strength and the line profile must be calculated for every relevant electronic, rotational, and vibrational transition. The input data for these calculations are provided by spectroscopic databases such as HITRAN \cite{Gordon2022JQSRT.27707949G}, HITEMP \cite{Rothman2010JQSRT.111.2139R}, and ExoMol \cite{Tennyson2016JMoSp.327...73T}.

In recent years, primarily through the efforts of the ExoMol team, the amount of available spectroscopic data has expanded significantly. This growth includes both the number of astrophysically relevant molecules and the quantity of recorded transitions. For instance, the latest line lists for species like methane (\ch{CH4}) and vanadium oxide (VO) now contain on the order of $10^{11}$ individual transitions. Consequently, computing opacities as a function of temperature, pressure, and wavelength has become an extremely demanding computational task.

To address this challenge, the open-source opacity calculator HELIOS-K\footnote{\url{https://github.com/exoclime/HELIOS-K}} was developed by \cite{Grimm2015ApJ...808..182G} to efficiently compute absorption coefficients for molecules, atoms, and ions. HELIOS-K is designed to run on graphics processing units (GPUs) rather than traditional CPU cores. This not only increases computational efficiency but also enables thousands of calculations to be performed in parallel on a single GPU. This high level of performance is essential for processing the massive line lists provided by databases like ExoMol within a reasonable timeframe.

Using HELIOS-K, an extensive database of absorption coefficients was calculated and tabulated across a broad temperature-pressure space for nearly all species available in the ExoMol and HITEMP databases \cite{Grimm2021ApJS..253...30G}. This opacity grid is publicly accessible via the DACE platform of the NCCR PlanetS\footnote{\url{https://dace.unige.ch}}, either through a web interface or a Python script. It represents one of the largest collections of high-resolution absorption coefficients relevant to (exo)planetary science and is widely used by the exoplanet atmosphere research community (e.g. \cite{Line2021Natur.598..580L,Keles2022MNRAS.513.1544K,Hu2024Natur.630..609H,August2025A&A...695A.171A,Coulombe2025AJ....170..226C}).

\subsection{Atmospheric chemistry}

The opacities that enter the radiative transfer equation require knowledge of the atmosphere's chemical composition, i.e. the number densities $n$ of all atmospheric constituents. Besides radiative transfer, chemistry also directly affects the basic atmospheric structure and dynamics via the mean molecular weight. Furthermore, the chemistry of condensing species is linked to cloud formation descriptions that, in turn, also again impact the radiative transfer calculations. The chemical composition can in principle be determined by either a priori assuming a composition, by using an equilibrium chemistry, or by employing a detailed (photo)chemical-kinetic network. The latter approaches require knowledge of the relevant thermochemical data as well as fundamental parameters of the modelled system, such as its elemental abundances.

\subsubsection{Chemical equilibrium}

Chemical equilibrium is a simplifying assumption in which the entire chemical system relaxes to a stationary state that minimizes the Gibbs free energy. This equilibrium state is unique and independent of the system’s initial conditions. Calculating the chemical composition under the assumption of equilibrium involves either directly minimizing the Gibbs free energy to find the number densities of all chemical species or solving a system of equations based on the law of mass action combined with element conservation. Mathematically, the former is a global optimization problem, while the latter forms a nonlinear system of equations.

For very simple systems, these problems can be solved analytically (e.g. \cite{Heng2016ApJ...816...96H, Heng2016ApJ...829..104H}). However, more complex scenarios typically involve dozens of elements and hundreds of gas-phase species and condensates, making analytic solutions impractical. In such cases, a full numerical solution must be computed.

The computation of chemical equilibrium solutions has a long history in the modeling of brown dwarf and planetary atmospheres (e.g., [29]). Several established computer codes with varying levels of complexity and different intended applications are available. Examples include the NASA CEA code \cite{Gordon1994}, CONDOR \cite{Lodders1993E&PSL.117..125L}, \textsc{GGChem} \cite{Woitke2018A&A...614A...1W}, or \textsc{FastChem} \cite{Stock2018MNRAS.479..865S, Stock2022MNRAS.517.4070S, Kitzmann2024MNRAS.527.7263K}.

\textsc{FastChem}, in particular, is one of the fastest chemical equilibrium codes developed and is available as open-source software\footnote{\url{https://github.com/NewStrangeWorlds/FastChem}}. It employs the law of mass action approach but splits the multidimensional, nonlinear system of equations into a set of one-dimensional problems that are solved iteratively. In many cases, these one-dimensional equations can be solved analytically, leading to a significant increase in computational efficiency.

\subsubsection{Non-equilibrium chemistry}
\label{sec:neqchem}

While chemical equilibrium provides substantial advantages for atmospheric modeling, it is important to recognize that not all atmospheric processes can be accurately described under equilibrium assumptions. In  environments, where chemical reactions do not have sufficient time to reach equilibrium, kinetic or photochemical modeling becomes necessary to accurately capture chemical behavior.

In general, the change in a species’ number density ($n_i$) over time is determined by source ($\mathcal P$) and loss ($\mathcal L$), as well as transport fluxes ($\phi$) in the horizontal and vertical directions:
\begin{equation}
    \frac{\partial n_i}{\partial t} = \mathcal{P}_i - \mathcal{L}_i + \nabla \phi_i \ .
\end{equation}

Production and loss processes are typically described by a network of chemical kinetic and photochemical reactions. To reliably calculate the abundance of a given chemical species, all relevant reactions involving it must be known. This makes non-equilibrium chemistry calculations significantly more complex than the simplifying assumption of chemical equilibrium.

Many chemical kinetics codes are proprietary (e.g. \cite{Venot2012A&A...546A..43V}, \cite{Hu2012ApJ...761..166H}, \cite{Moses2011ApJ...737...15M}, \cite{Allen1981JGR....86.3617A}, \cite{Rimmer2016ApJS..224....9R}) or limited to specific planet types, such as terrestrial planets (e.g. \cite{Segura2003AsBio...3..689S}). The non-equilibrium chemistry code VULCAN \cite{Tsai2017ApJS..228...20T} was one of the first publicly available codes\footnote{\url{https://github.com/exoclime/VULCAN}}  dedicated to modeling atmospheric chemistry in exoplanets.
VULCAN's original Carbon-Hydrogen-Oxygen-Nitrogen (CHON) network was based extensively on the NIST database \cite{NIST} with some additions from other sources, such as \cite{Moses2011ApJ...737...15M}, and has since been expanded to include photochemistry \cite{Tsai2021ApJ...923..264T} and additional chemical species of interest to the exoplanet scientific community such as sulphur \cite{Tsai2023Natur.617..483T} and phosphorous chemistry \cite{Lee2024ApJ...976..231L}.

Chemical reaction networks can become quite large, connecting hundreds of species through sometimes thousands of reactions. This makes numerical solutions challenging and often computationally expensive. However, it is often unnecessary to include all possible reactions, as many may be too slow or unimportant to significantly affect species’ abundances. To address this, \cite{Tsai2018ApJ...862...31T} replaced VULCAN’s full reaction network with a chemical relaxation scheme that depends on only a handful of independent source and sink terms. The chemical timescales were determined using a pathway analysis tool that identifies the rate-limiting reactions as a function of temperature and pressure. This relaxation scheme compares well with results from the full network but can be solved much faster.

Nevertheless, even with these improvements, VULCAN remained too slow to be integrated within general circulation models (GCMs) that study the effects of hydrodynamical transport and diffusion on chemical abundances. To overcome this, \cite{Lee2023A&A...672A.110L} developed the \textsc{Mini-chem} code as an offshoot of VULCAN, available as an open-source software package\footnote{ \url{https://github.com/ELeeAstro/mini\_chem} }. \textsc{Mini-chem} employs the method introduced in \cite{Tsai2022A&A...664A..82T}, which uses the concept of net reactions to significantly reduce the size of the chemical network. This innovation finally made it possible to efficiently model non-equilibrium chemistry within a GCM \cite{Lee2023A&A...672A.110L,Lee2023MNRAS.523.4477L}.

\subsection{Cloud microphysics}

A fundamental part of atmospheres which has a significant and complex effect on the climate, dynamics, energy balance and observational properties is the formation and dynamics of clouds. 
For hot gas giant exoplanets and brown dwarfs above the L-T transition, mineral refractory materials such as rutile (\ch{TiO2}), corundum (\ch{Al2O3}) and iron (Fe) can form, as well as silicate materials such as enstatite (\ch{MgSiO3}), forsterite (\ch{Mg2SiO4}) and quartz (\ch{SiO2}).
For cooler planets and brown dwarfs, sulphides such as alabandite (MnS), sodium sulfide (Na$_{2}$S) and zinc sulfate (ZnS) could potentially form, as well as potassium salt cloud composed of potassium chloride (KCl) \cite{Morley2012ApJ...756..172M}.
For the coldest Y dwarfs and exoplanets, the familiar water (\ch{H2O}) clouds can form as well as ammonia (\ch{NH3}) and \ch{CH4}, similar to our Solar System gas giants.
The only truly cloud free exoplanet atmospheres are ultra-hot Jupiters with nightside temperatures above $\sim$1500\,K, only a small part of the exoplanet population, which suggests clouds are ubiquitous feature of atmospheres across the exoplanet parameter range.

Due to their major importance and complex feedback mechanisms, developing understanding of cloud formation microphysical processes is a highly important topic to tackle, but is rarely considered in detail by the field.
However, recently, significant progress has been made developing complex cloud microphysical models for exoplanet atmospheres \cite{Gao2021JGRE..12606655G}.
For example, in \cite{Lee2023MNRAS.524.2918L}, the `mini-cloud' time dependent cloud formation model was developed, including nucleation and condensation/evaporation of mineral materials.
This used the `moment/bulk method', originally developed in the context of asymptotic giant branch stars \cite{Gail1988A&A...206..153G}, to evolve cloud properties such as particle size and number density in an numerically efficient way, enabling complex cloud microphysics to be included in large scale atmospheric simulation platforms such as GCMs.
Furthermore, in \cite{Lee2025A&A...695A.111L} a microphysical theory was developed including nucleation, condensation/evaporation and collisional growth processes.
Due to the addition of collisional growth, this model represents a very complete microphysical model and large step in microphysical understanding of mineral clouds. 

The cloud particle size distribution plays a pivotal role in determining cloud radiative feedback and, consequently, the observable properties of planetary atmospheres. While so-called bin or spectral methods exist to directly compute the size distribution \cite{Gao2018ApJ...855...86G, Powell2019ApJ...887..170P}, these highly detailed frameworks are computationally expensive, making them impractical for use in complex GCMs. To address this limitation, \cite{Lee2025A&A...698A.220L} recently extended the mass moment method to account for the effects of a polydisperse size distribution on cloud microphysical properties, finding significant differences compared to previous studies that assumed monodisperse particle sizes (e.g. \cite{Helling2008A&A...485..547H}).

\subsection{Climate}

As stated above, solving the full system shown in Figure \ref{fig:atmosphere_flowchart} in full generality is often not possible. This would not only need an excessive amount of computational power but would also require physical or chemical input data that is rarely available over the entire parameter space one intends to model. This, for example, refers to incomplete line lists of molecules or unreliable chemical reaction networks to describe cloud formation in detail. Therefore, atmospheric models use simplifications of the full theoretical descriptions or make additional assumptions about to system to make the problem solvable. The level of detail specific atmospheric models use is also often defined by the available observational data. If, for example, only a single photometric measurement of an exoplanet is available, it usually makes little sense to simulate its atmosphere with a full three-dimensional general circulation model including non-equilibrium chemistry and a detailed cloud-formation scheme.

\subsubsection{One-dimensional models}
\label{sec:1d_climate}

One central simplification that is very often done is to assume that the atmosphere can be described by a one-dimensional model, thereby neglecting the full three-dimensional nature of the planets. Such an atmospheric model then only describes a somewhat global average state of the atmosphere, for example, an average over the day or nightside of the planet.. However, given that we often only have observations that do not resolve the full 3D information of a planet, such a simplification can be justified. 
In contrast to fully three-dimensional models, 1D models have significantly lower computational demands. This allows them to incorporate more complex physical and chemical processes, such as detailed chemical kinetics and cloud formation treatments as well as high-resolution line-by-line radiative transfer calculations.

One-dimensional models have a long history in atmospheric modelling that is not only limited to planetary atmospheres. Stellar atmospheres and brown dwarfs are also, for example, very often modelled assuming a 1D atmospheric structure \cite{Hauschildt1997ApJ...483..390H, Ackerman2001ApJ...556..872A}. Planetary science also in past used 1D models to theoretically model atmospheres. This includes both Solar System planets as well as exoplanets (e.g. \cite{Kasting1988Icar...74..472K}, \cite{Kasting1993Icar..101..108K}, \cite{Molliere2019A&A...627A..67M}, \cite{Goyal2018MNRAS.474.5158G}). 

A general atmospheric model, specifically designed for exoplanets and developed in the frame of the NCCR PlanetS is HELIOS (\cite{Malik2017AJ....153...56M}). HELIOS is an open-source model\footnote{\url{https://github.com/exoclime/HELIOS}} that treats the atmosphere as a one-dimensional, plane-parallel system. It, thus, does not has to solve the 3D fluid dynamics equations but rather only assumes hydrostatic equilibrium. Initially only limited to irradiated  and self-luminous atmospheres of gaseous planets (\cite{Malik2019AJ....157..170M}), it was later adapted to also treat atmospheres of terrestrial planets (\cite{Malik2019ApJ...886..142M}). 

HELIOS has been successfully employed to model a broad range of exoplanet atmospheres and brown dwarfs (e.g. \cite{Crossfield2022ApJ...937L..17C}, \cite{Deline2022A&A...659A..74D}, \cite{Wong2021AJ....162..256W}). It has also been used to generate self-consistent atmospheric models for training machine-learning atmospheric retrieval approaches (e.g. \cite{Oreshenko2020AJ....159....6O, Lueber2023ApJ...954...22L}). Furthermore, it was also coupled to a planet interior model to consistently model a mini-Neptune exoplanet from its interior to the upper atmosphere in \cite{Guzman2022MNRAS.513.4015G}.  

One particularly interesting application of HELIOS in combination with the \textsc{FastChem} chemistry code was its use in studying the atmospheric structure and chemistry of the ultra-hot Jupiter KELT-9b, which, with an estimated equilibrium temperature in excess of 4000\,K, remains the hottest exoplanet discovered to date \cite{Kitzmann2018ApJ...863..183K}. The model predicted the strong presence of the hydrogen anion (\ch{H-}) and an atmosphere largely devoid of molecules due to extensive thermal dissociation. Instead of molecules, the atmosphere was found to be dominated by atoms and ions, including many refractory elements such as iron and titanium. The presence of these species was confirmed shortly afterward through high-resolution spectroscopic observations (see Section \ref{sec:characterization_high_res}).

\subsubsection{General-circulation models}

General Circulation Models (GCMs) are powerful tools used to study the atmospheric dynamics and climate of extrasolar planets. Contemporary GCM simulations are essential for understanding the complex global feedback interactions that shape exoplanetary atmospheres. GCMs simulate large-scale circulation patterns, heat transport, and radiative processes that together determine a planet’s climate. Able to be coupled with chemical and cloud models, they are able to uncover and give understanding of the processes that give rise to the observed properties of exoplanet atmospheres.

GCMs are models that solve various flavours of the Navier-Stokes equations across a rotating 3D spherical (or spherical-like) computational grid \cite{Mayne2014A&A...561A...1M}. A key component of any GCM is the dynamical core, which solves these equations to simulate large-scale circulation features, such as jet streams and Hadley cells. 

The dynamical core is responsible for modeling the movement of air masses and the transport of heat and momentum throughout the atmosphere. Many GCMs employed for exoplanet and brown dwarf atmospheres have been adapted from those developed for Earth and other Solar System planets. The seminal hot Jupiter GCM simulations performed by Showman (\cite{Showman2008ApJ...682..559S, Showman2009ApJ...699..564S}) successfully predicted an eastward shift the dayside hotspot from the sub-stellar point due to the formation of strong equatorial jets. These jets and its effect on energy transport and balance then impact the overall thermal phase curve and dayside/nightside thermal contrast.

In addition to the dynamical core, GCMs include various components to describe processes such as radiative transfer, the temperature structure computations, and atmospheric chemical composition. Processes that occur on scales smaller than the model resolution, and therefore cannot be explicitly resolved, are typically represented in a parameterized manner. Examples include cloud formation, precipitation, and convection. These parameterizations are often calibrated using observational data or higher-resolution simulations to enhance model accuracy.

We highlight two GCMs used in NCCR PlanetS research, THOR and Exo-FMS.
THOR is the first open-source GCM developed from scratch to explicitly model the atmospheres of exoplanets\footnote{\url{https://github.com/exoclime/THOR}} \cite{Mendonca2016ApJ...829..115M, Deitrick2022MNRAS.512.3759D}. Running on GPUs for computational performance, it solves the non-hydrostatic Euler equations on an icosahedral grid.
Exo-FMS, developed at the University of Oxford, UK. is an offshoot of GFDL's Earth climate model, The Flexible Modelling System (FMS), modified for exoplanet atmospheres \cite{Hammond2017ApJ...849..152H,Lee2021MNRAS.506.2695L}.
Exo-FMS solves the primitive equations of meteorology across a cubed-sphere grid.

Both THOR and Exo-FMS have investigated the transport of chemical species, with THOR utilising the paramterised chemical relaxation method \cite{Tsai2018ApJ...862...31T,Mendonca2018ApJ...869..107M} to investigate the 3D chemical composition of WASP-43b, a prime JWST target.
Exo-FMS uses the \textsc{Mini-chem} kinetic chemistry methodology (Section \ref{sec:neqchem}) to more self-consistently model the non-equilibrium vertical, horizontal and meridional mixing in the atmosphere.
Exo-FMS is also able to feedback the changing chemical species into its radiative-transfer scheme.

Some of the most sophisticated contemporary global atmosphere simulations were recently performed by Exo-FMS, which coupled kinetic chemistry, convection and cloud formation with consistent radiative-transfer feedback to investigate effects of cloud and chemistry connections on spectral variability in brown dwarf atmospheres \cite{Lee2023MNRAS.523.4477L,Lee2024MNRAS.529.2686L} and for WASP-43b as part of the NIRSpec GTO programme \cite{Challener2024ApJ...969L..32C}.
This has been recently been extended to include detailed cloud microphysics, investigating the spectral variability of the prime Y-dwarf JWST target WISE 0359-54 \cite{Lee2025A&A...695A.111L}.

\section{Atmospheric Retrieval Frameworks}
 
\begin{figure}[t]
\centering
\includegraphics[width=0.8\textwidth]{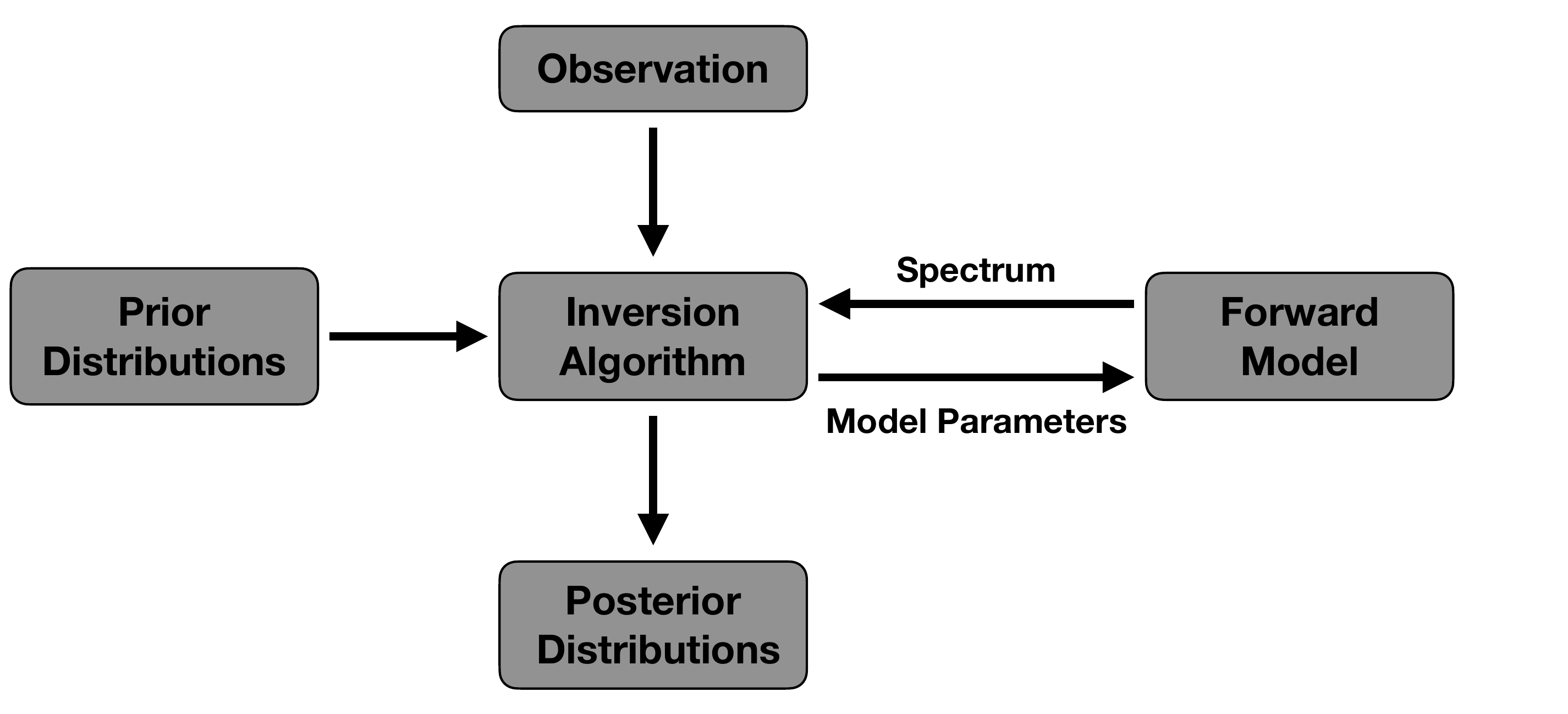}
\caption{Essential components of a retrieval framework for atmospheric characterization of brown dwarf or exoplanet observations.}
\label{fig:retrieval_flowchart}
\end{figure}

The atmospheric models discussed in the previous section aim to predict the state of an atmosphere based on a based on a theoretical model involving descriptions for the various chemical and physical processes that govern the atmosphere. While self-consistent models are invaluable for investigating detailed atmospheric processes, they have limitations when it comes to robustly interpreting observed spectra. By definition, self-consistent models assume that these physical and chemical processes are already known. However, the novelty and diversity of exoplanetary atmospheres mean that we often have very limited prior knowledge. As a result, real exoplanet atmospheres may differ significantly from the assumptions and approximations built into self-consistent models.

In contrast, an atmospheric \textit{retrieval framework} addresses the inverse problem: it uses observational data, such as a transmission spectrum of an exoplanet, to infer the atmospheric properties that produced the observed signal. These inferred properties include key atmospheric characteristics such as chemical composition, temperature structure, and cloud coverage. Unlike traditional self-consistent models, which predict atmospheric structures and spectra from a limited set of fundamental parameters, retrieval methods treat a wide range of atmospheric properties and processes as free parameters. This approach enables observational data to directly constrain the atmospheric composition and structure. 

A conceptual overview of this method is shown in Figure \ref{fig:retrieval_flowchart}. A typical retrieval framework consists of two main components: the \textit{forward model}, which computes a theoretical spectrum based on a given set of atmospheric parameters and the \textit{inversion algorithm}, which determines the posterior distributions of those parameters that best fit the observed spectrum, incorporating prior knowledge of the system.

With the increase in exoplanet and brown dwarf observations, a wide range of retrieval frameworks have been developed over the past decade. This includes NEMESIS \cite{Irwin2008JQSRT.109.1136I, Lee2012MNRAS.420..170L}, CHIMERA \cite{Line_2013}, TAU-REX \cite{Waldmann2015ApJ...802..107W}, HELIOS-r \& HELIOS-r2 \cite{Lavie2017AJ....154...91L, Kitzmann2020ApJ...890..174K}, HELA \cite{Marquez-Neila2018NatAs...2..719M}, petitRadTrans \cite{Molliere2019A&A...627A..67M}, AURORA \cite{Welbanks2021ApJ...913..114W}, or POSEIDON \cite{ MacDonald2023JOSS....8.4873M}, to name a few\footnote{See \url{https://arxiv.org/pdf/2303.12925} for a more comprehensible list of retrieval codes.}. Many of these code have been released as open-source software. 

Under the NCCR PlanetS framework, two open-source retrieval codes, HELIOS-r and HELIOS-r2 \cite{Lavie2017AJ....154...91L, Kitzmann2020ApJ...890..174K}, were developed. The original HELIOS-r code was designed specifically for emission spectroscopy of self-luminous objects. In contrast, HELIOS-r2\footnote{HELIOS-r2 has been recently renamed to BeAR: \url{https://github.com/NewStrangeWorlds/BeAR}.} is a more versatile tool capable of characterizing transmission, emission, and secondary-eclipse spectra of both brown dwarfs and exoplanets. Like many other codes discussed in this chapter, HELIOS-r2 performs most of its computations on a GPU, significantly enhancing its computational efficiency. It has been widely applied to the study of brown dwarfs \cite{Kitzmann2020ApJ...890..174K, Lueber2023ApJ...954...22L, Lueber2022ApJ...930..136L} and exoplanets, ranging from hot gas giants \cite{Bourrier2020A&A...637A..36B, Petit2024A&A...692A..83P, Powell2024Natur.626..979P} to rocky terrestrial planets \cite{Patel2024A&A...690A.159P}.

\subsection{Forward model}

A retrieval framework typically includes a simplified atmospheric model, commonly referred to as a \textit{forward model}, which generates a theoretical spectrum based on a given set of parameters. A central component of a forward model is therefore a radiative transfer scheme that transforms the atmospheric structure and chemical composition into a spectrum that can be compared to the actual observation. The complexity of these forward models can vary widely. For example, some models may assume that the abundance of a given species is constant with altitude, or that the temperature follows a fixed, predefined profile \cite{Madhusudhan2009ApJ...707...24M}, characterized by a limited number of free parameters. Others may allow both temperature profiles and molecular abundances to vary more freely throughout the atmosphere.

Since a forward model in a classical retrieval framework is evaluated tens of thousands of times or much more, it must be highly efficient to minimize computation time. This computational constraint typically precludes the use of detailed atmospheric physics or chemistry, as such complexity would make the retrieval process prohibitively expensive. As a result, many atmospheric processes are included in a simplified, parameterized form.

Unlike self-consistent atmospheric models, forward models used in retrieval frameworks do not necessarily enforce physical constraints such as radiative equilibrium or the conservation of elemental abundances. As a result, while they offer greater flexibility in fitting observational data, they may sometimes produce non-physical or internally inconsistent atmospheric states. Nevertheless, by relaxing the requirement of self-consistency, these models can explore a much broader range of atmospheric conditions, often achieving a closer match to the observed spectra than self-consistent models. The latter, however, can still serve as important tools for benchmarking the more simplified forward models commonly used in atmospheric retrievals. They allow, for example, to test the validity of specific modelling assumptions and can be used to derive constraints on prior distributions.

\subsection{Inversion methods} 

Retrieval models employ a range of techniques to solve the inverse problem, that is, to determine the posterior distribution of a set of free model parameters based on observational data. These techniques include Markov Chain Monte Carlo (MCMC) methods, Bayesian inference, and increasingly, machine learning approaches.

Inversion algorithms rely on prior distributions to describe existing knowledge (or assumptions) about the parameters. For example, if a planet’s mass has been measured via radial velocity, one might use a Gaussian prior to describe its surface gravity. Conversely, if no prior information is available for a given parameter, an uninformed or uniform prior may be used to allow the algorithm to explore a broader parameter space. Using these priors, the inversion technique searches the parameter space to derive the posterior distributions.

Among the most commonly used inversion techniques are MCMC methods and Bayesian inference frameworks. Within Bayesian approaches, nested sampling \cite{Skilling2004AIPC..735..395S} has gained popularity as a robust method for estimating posterior distributions. Several implementations of nested sampling are available \cite{Feroz2008MNRAS.384..449F, Brewer2009arXiv0912.2380B, Higson2019S&C....29..891H}. Another Bayesian algorithm often applied to solar system observations is Optimal Estimation (OE). OE is significantly faster than nested sampling, but it assumes Gaussian priors and is therefore best suited to cases where prior knowledge is sufficiently well constrained.

Inversion algorithms based on Bayesian statistics offer a significant advantage: they enable the calculation of the Bayesian evidence, which allows for the computation of the Bayes factor between competing models. The Bayes factor quantitatively compares how well one model explains the observations relative to another. In this way, it serves as a formal, quantitative expression of Occam’s razor, favoring simpler models unless a more complex one provides a substantially better explanation of the data \cite{Trotta2008ConPh..49...71T}. 

A persistent challenge in atmospheric retrieval is the trade-off between physical and chemical accuracy versus computational efficiency. In practice, forward models, which convert chemical abundances and opacities into theoretical spectra, must often be simplified to ensure reasonable computation times. To address this, machine learning (ML) techniques are emerging as a promising alternative, offering rapid and efficient characterization of exoplanetary atmospheres. Deep learning architectures, in particular, are capable of modeling the high-dimensional, non-linear relationships between planetary parameters and observed spectra.

Several ML approaches have been applied to the analysis of exoplanet and brown dwarf spectra, including deep belief networks \cite{Waldmann2016ApJ...820..107W}, random forests \cite{Marquez-Neila2018NatAs...2..719M, Oreshenko2020AJ....159....6O}, generative adversarial networks \cite{Zingales2018AJ....156..268Z}, convolutional neural networks \cite{Himes2022PSJ.....3...91H}, and Bayesian neural networks \cite{Cobb2019AJ....158...33C}. More recent advancements include simulation-based inference methods that can perform Bayesian model comparisons, akin to classical nested sampling \cite{Cranmer2020PNAS..11730055C, Vasist2023A&A...672A.147V}.

Despite their promise, ML models introduce new sources of uncertainty. These include limitations in the training data, sensitivity to hyperparameter tuning, and interpretability challenges. Thus, despite their predictive power, deep learning models are often regarded as “black boxes” that hide the relevant physics and are not easily interpretable. In the exoplanet research field, there is also a lack of large labeled datasets for training, which means that ML models are almost exclusively trained on synthetic data derived from self-consistent or parametrized forward models. Nevertheless, machine learning approaches enable the inclusion of significantly more complex forward models in retrieval analyses than is feasible with traditional methods. This was demonstrated, for example, by \cite{GuzmanMesa2022MNRAS.513.4015G}, who used a self-consistently coupled interior–atmosphere model to train a random forest and perform a retrieval analysis of the sub-Neptune GJ 436 b. This allowed for a joint retrieval of both the interior structure and the planet's atmosphere. Incorporating such a complex model into retrievals using techniques like MCMC or nested sampling would be computationally prohibitive, as the time required for convergence would be unmanageable.

\section{Observational characterization}

Space observatories offer a wide range of observational capabilities in terms of wavelength coverage and spectral resolution. Simpler missions, such as CHEOPS, Kepler, and TESS, operate with a single photometric bandpass in the optical to near-infrared range. While limited in spectral detail, these missions typically allow for longer observational baselines per target, enabling the observation of long-term phase curves (e.g. \cite{Meier2023A&A...677A.112M}). More details on observations with CHEOPS can be found in Chapter 10. The Hubble Space Telescope (HST) and James Webb Space Telescope (JWST), on the other hand, are equipped with spectrographs that offer broad wavelength coverage. HST enables observations across the ultraviolet, visible, and near-infrared regions, while JWST covers the near-infrared to mid-infrared wavelengths. However, due to their compact design and wide wavelength coverage, these space-based instruments typically have limited spectral resolution. As a result, they can detect broad molecular absorption features in exoplanet spectra, but are often unable to resolve individual spectral lines. 

In contrast, high-resolution observations of individual spectral lines are predominantly conducted using ground-based telescopes. Although these instruments do not benefit from the absolute stability of the space environment, their high spectral resolution allows access to additional observables. For example, Doppler shifts caused by atmospheric winds can displace or distort spectral lines. Moreover, sharp absorption lines may stand out above the broad, cloud-induced continuum, enabling high-resolution spectrographs to probe atmospheres that are otherwise challenging to analyze with space-based instruments.

A key method used in analyzing exoplanet spectra is the \textit{cross-correlation technique} \cite{Snellen2010Natur.465.1049S}. This approach requires high spectral resolution, such as that provided by instruments like HARPS, ESPRESSO, CRIRES, and the upcoming ANDES and METIS on the Extremely Large Telescope (ELT), along with precise knowledge of the positions and relative strengths of the species' line transitions. More details on high-resolution spectroscopy can be found in Chapter 12.

Over the past decade, atmospheric retrieval techniques have become central to the characterization of exoplanet and brown dwarf atmospheres. Traditionally, brown dwarf spectra were interpreted by comparing observations with pre-computed grids of self-consistent atmospheric models. The best-fitting model from these grids would then provide insights into the object's mass, evolutionary state, atmospheric composition, and temperature structure (e.g., \cite{Burgasser2006ApJ...637.1067B, Burrows2006ApJ...640.1063B}). Early exoplanet retrieval studies adopted a similar grid-based approach \cite{Madhusudhan2009ApJ...707...24M}.

Since those initial efforts, a variety of more advanced statistical tools have been developed, ranging from Bayesian inference frameworks to machine learning techniques. Many of the retrieval methods used in exoplanet and brown dwarf studies were originally developed in the context of solar system remote sensing, including optimal estimation methods \cite{Irwin2008JQSRT.109.1136I}.

\subsection{Space-based observations}

\subparagraph{Hubble Space Telescope}

The Hubble Space Telescope (HST) has played a pivotal role in advancing our understanding of exoplanet atmospheres through numerous key discoveries. It has been instrumental in atmospheric characterization, particularly via transit spectroscopy and eclipse observations. HST provides transmission spectra in the 0.2–1.7 $\mu$m range using the Wide Field Camera 3 (WFC3), which operates with both UVIS (0.2–1.0 $\mu$m) and NIR (0.85 - 1.7 $\mu$m) channels. Additionally, the Space Telescope Imaging Spectrograph (STIS) extends HST’s capabilities into the ultraviolet and optical range, covering wavelengths from about 0.1 to 1 $\mu$m.

HST’s near-infrared observations, especially through WFC3, have enabled the detection of molecular species in the atmospheres of hot Jupiters and other gas giants. The WFC3 spectral range contains strong absorption features from \ch{H2O}, along with contributions from molecules such as \ch{CH4}, \ch{NH3}, and hydrogen cyanide (HCN). Transmission spectra obtained with WFC3 have been widely used to characterize exoplanet atmospheres. A notable early example is the observation of the super-Earth GJ 1214b \cite{Berta2012ApJ...747...35B}, which revealed a featureless spectrum. Over the past decade, most WFC3 analyses have focused on detecting water or methane features near 1.4 $\mu$m \cite{Deming2013ApJ...774...95D, Wakeford2013MNRAS.435.3481W, Kreidberg2014ApJ...793L..27K}.

STIS has further enhanced atmospheric studies by providing optical and near-ultraviolet coverage. This is particularly important for detecting alkali metals such as sodium (Na) and potassium (K), as well as molecules with strong short-wavelength opacities like titanium oxide (TiO) and vanadium oxide (VO) \cite{Allen2024AJ....168..111A}. These optical measurements complement WFC3 data, expanding the spectral baseline available for atmospheric retrievals. HST has, for example, been used to study the atmospheres of ultra-hot Jupiters, where temperatures exceed 2000\,K. For instance, in WASP-76b, HST detected TiO and \ch{H2O} in the transmission spectrum. Its secondary eclipse spectrum revealed a strong CO emission feature, suggesting a temperature inversion in the planet’s atmosphere \cite{Fu2021AJ....162..108F}. These findings underscore the complexity of chemical and thermal processes in extreme planetary environments.

In addition, HST observations have provided insight into clouds and hazes. Several studies have analyzed the impact of aerosols on transmission spectra \cite{Gibson2012MNRAS.422..753G, Kreidberg2014ApJ...793L..27K, Kreidberg2014Natur.505...69K, Sing2016Natur.529...59S}. STIS, in particular, is sensitive to optical slopes in transmission spectra, which help constrain scattering processes such as molecular Rayleigh scattering or the presence of high-altitude aerosols \cite{Pont2008MNRAS.385..109P, Pont2013MNRAS.432.2917P, Wakeford2015A&A...573A.122W, Pinhas2017MNRAS.471.4355P}

As the number of transmission spectra from WFC3 has grown, statistical studies have emerged to explore trends across different planetary systems. These studies investigate correlations between atmospheric properties -- such as metallicity, water abundance, cloud coverage, and temperature -- and planetary characteristics like mass and radius \cite{Sing2016Natur.529...59S, Barstow2017ApJ...834...50B, MacDonald2017ApJ...850L..15M, Fisher2019ApJ...881...25F, Tsiaras2018AJ....155..156T, Welbanks2019ApJ...887L..20W}. In addition to exoplanet observations, the HST has also provided high-quality data on brown dwarfs, offering valuable insights into their cloud properties and variability (e.g. \cite{Apai2013ApJ...768..121A, Lueber2024A&A...690A.357L}).

While HST has been foundational in exoplanet atmosphere research, it also faces limitations, particularly in its narrower spectral range compared to more advanced observatories like the James Webb Space Telescope (JWST). With its broader spectral coverage and higher precision, JWST is expected to surpass many of HST’s capabilities, enabling more detailed and accurate atmospheric characterizations. Nonetheless, HST’s extensive archival data remains an invaluable resource for comparative studies and for refining techniques that will inform future observations \cite{Fisher2024MNRAS.535...27F}. Moreover, HST remains the only space-based observatory currently capable of performing ultraviolet observations, a wavelength range not accessible to JWST. This ensures HST’s continued relevance and scientific value in the era of next-generation telescopes.

\subparagraph{James Webb Space Telescope}

The advent of data from the James Webb Space Telescope (JWST) has dramatically improved the data quality and wavelength coverage of exoplanet atmospheric spectra. JWST’s four spectroscopic instruments cover wavelengths from 0.6 $\mu$m to 28 $\mu$m, providing an unprecedented wealth of information. This broader wavelength range has already led to the discovery of new molecules and the confirmation of previous detections \cite{bell23}.

Notably, JWST has provided the first detections of \ch{CO2}, \ch{H2S}, or \ch{SO2} in an exoplanet atmosphere \cite{ERS2023Natur.614..649J,Fu2024Natur.632..752F, 2024Natur.626..979P}, allowing for a more comprehensive study of atmospheric chemistry and physics. JWST will also shed light on the interior structures and compositions of sub-Neptunes, which, despite being among the most common exoplanetary bodies, are absent from our Solar System. This includes, for example the sub-Neptune K2-18 b, where \ch{CH4}, \ch{CO2}, and \ch{H2O} have been found \cite{Madhusudhan2023ApJ...956L..13M}. 

Beyond gas-phase composition, JWST data has greatly enhanced our understanding of cloudy exoplanets \cite{kempton23,grant23, Inglis2024ApJ...973L..41I} and expanded the range of smaller planets that can be observed \cite{lustig-yaeger23,moran23,kirk24, Rathcke2025ApJ...979L..19R}. 

JWST has also significantly improved the quality of brown dwarf observations, surpassing previous capabilities of the Spitzer and Hubble space telescopes. This was, for example, demonstrated through the JWST Early Release Science Program (ERS) \cite{2022PASP_Hinkley}), which produced the first higher-resolution spectrum of VHS 1256 b, spanning from 0.9 to 18 $\mu$m \cite{Miles2023ApJ...946L...6M}.

These observations have revealed silicate cloud features at 10 $\mu$m, signs of \ch{CO2} gas, and simultaneous detections of both CO and \ch{CH4}, suggesting disequilibrium chemistry driven by strong vertical mixing. Some brown dwarf spectra are already of sufficient quality to detect temporal variability \cite{Apai2013ApJ...768..121A}, often attributed to horizontal variations in cloud optical thickness, high-altitude hazes, or patchy clouds \cite{Buenzli2014ApJ...782...77B, Yang2015ApJ...798L..13Y}. Notably, \cite{Vos2023ApJ...944..138V}) used JWST spectra to retrieve the inhomogeneous atmospheres of two highly variable early T-type brown dwarfs.

Across JWST General Observer (GO) Cycles 1-3 and several Guaranteed Time Observation (GTO) programs, over 150 exoplanets have already been observed, with further data expected in upcoming cycles. In addition, more than 150 brown dwarfs have been observed as well, covering all spectral classes from L0 to Y4.

Many of the modeling and retrieval frameworks developed within PlanetS have made significant contributions to the interpretation of JWST data. This includes, for example, the detection and explanation of \ch{SO2} in the atmosphere of WASP-39 b \cite{Tsai2023Natur.617..483T, 2024Natur.626..979P}, as well as efforts to understand the variability of the super-Earth 55 Cnc e \cite{Patel2024A&A...690A.159P}.

\subsection{Characterizing exoplanet atmospheres at high spectral resolution}
\label{sec:characterization_high_res}

\subparagraph{Detection of species via the cross-correlation technique}

Unlike molecules, atoms and ions do not produce strong, broad absorption bands that are easily detectable with low- or medium-resolution spectroscopy. Instead, their spectral features consist of narrow lines that require high-resolution spectroscopy for detection. Currently, such resolution is only achievable with ground-based telescopes equipped with high-resolution spectrographs, though a limited number of narrow-band high-resolution modes are available with the STIS instrument on the HST (see, e.g., \cite{Sing2019AJ....158...91S}).

However, high spectral dispersion leads to increased photon noise per resolution element, making individual atomic or ionic absorption lines typically weaker than the noise. To address this, a technique known as the cross-correlation method is employed, which combines the contributions of many individual lines into one \textit{cross-correlation signal} \cite{Snellen2010Natur.465.1049S}. This technique uses model templates to identify the spectral fingerprints of specific species by cross-correlating the template with the observed spectrum and by summing over many spectral lines present in the data, thereby effectively averaging out photon noise.

In addition to enabling the detection of species with sparser spectral features than the broad molecular bands observable with space-based instruments, high-resolution cross-correlation spectroscopy offers several unique advantages. The method is highly robust against false positives because it relies on the distinctive wavelength patterns of many lines. Moreover, it can resolve sub-kilometer-per-second Doppler shifts, allowing for the direct measurement of hot Jupiter orbital velocities and inclinations v, as well as the detection of atmospheric rotation and wind patterns \cite{Snellen2010Natur.465.1049S,Louden2015}. This provides valuable and unique observables not accessible with low-resolution spectroscopy.

Application of high-resolution cross-correlation spectroscopy has increased strongly since the discovery of ultra-hot Jupiters post 2018. As discussed in Section \ref{sec:1d_climate}, the study by \cite{Kitzmann2018ApJ...863..183K} predicted that the transmission spectrum of the ultra-hot Jupiter KELT-9b should be dominated by absorption lines of atoms, due to the extreme temperatures that cause thermal dissociation of most molecules. This study suggested that observations of a single transit using a moderately large ground-based telescope could reveal the presence of atomic iron using the cross-correlation technique. This prediction was confirmed shortly thereafter through actual observations with the HARPS-N spectrograph on the TNG telescope in La Palma \cite{Hoeijmakers2018Natur.560..453H}. In addition, ionized iron and ionized titanium were also detected \cite{Hoeijmakers2018Natur.560..453H} although these were not expected \cite{Kitzmann2018ApJ...863..183K}, likely indicating that the upper atmosphere of the planet is significantly expanded. Subsequent transit observations of KELT-9b revealed even more species, including neutral sodium, as well as ionized chromium, scandium, and yttrium \cite{Hoeijmakers2019A&A...627A.165H}, and later, neutral calcium, vanadium, titanium, and chromium, along with ionized strontium, barium, and terbium \cite{Borsato2023A&A...673A.158B}.

These pioneering detections marked the beginning of a new era in exoplanet atmospheric studies: the systematic search for and identification of atomic, ionic, and molecular species using the cross-correlation technique applied to high-resolution, ground-based spectra in both transmission as well as day-side emission, and using these to place entirely new constraints on exoplanet atmospheres (e.g. \cite{Prinoth2022NatAs...6..449P,Stefan2023,Hoeijmakers2024,Seidel2025}).

The application of the cross-correlation technique to high-resolution spectra requires templates for the chemical species that one wants to detect in an exoplanet atmosphere. In order to perform cross-correlation calculations in a more general framework, \cite{Kitzmann2023A&A...669A.113K} published a publicly available library of standard cross-correlation templates and binary line masks. The templates are available for a large variety of atoms and ions, as well as six molecules and are intended for ultra-hot Jupiters. By using standard templates, the cross-correlation results of different observations can be compared in a systematic and quantitative way.

\subparagraph{Characterising ultra-hot Jupiters in 3D}

Ultra-hot Jupiters are among the most favorable targets for atmospheric characterization at high spectral resolution. Resolving their transmission spectra as a function of orbital phase offers a unique opportunity to probe the three-dimensional structure of their atmospheres. These planets are typically tidally locked, with a permanently irradiated dayside and a perpetually dark nightside, resulting in extreme temperature contrasts. Dayside temperatures often reach $\sim$3000\,K, while nightside temperatures can be more than 1000\,K lower.

This stark thermal gradient leads to distinct chemical regimes across the planet. On the intensely hot dayside, molecules, including \ch{H2}, undergo thermal dissociation. Many metals (e.g., Fe, Ti, V, Cr, Ni) exist in the gas phase, and only the most strongly bound molecules, such as \ch{H2O} and CO, survive in trace amounts. On the cooler nightside, a variety of metal oxides form and condense into clouds, as many are refractory with high condensation temperatures. This can lead to \textit{cold-trapping}, where condensed species become sequestered on the nightside and are globally depleted from the atmosphere \cite{Spiegel2009ApJ...699.1487S}.

High-resolution spectroscopic observations are sensitive to this 3D structure because the planet rotates significantly during a transit event. This means that transit observations scan a range of longitudes, leading to changes in the strength and radial velocity of the cross-correlation signal (see e.g. \cite{Ehrenreich2020,Prinoth2022NatAs...6..449P}). High-resolution transit observations therefore provide new insights into how chemical species and heat are distributed across the planet. However, fully exploiting the rich information in both high- and low-resolution spectra requires a deeper understanding of these three-dimensional processes. As a result, modeling efforts that bridge theory and observation are essential for interpreting current and future datasets.

One such approach involves post-processing detailed outputs from 3D general circulation models using the gCMRT radiative transfer tool \cite{Lee2022ApJ...929..180L} described in Section \ref{sec:radiative_transfer}. This enables the investigation of how a planet’s atmospheric structure and dynamics impact observable features. Ultra-hot Jupiters, in particular, show extreme spatial variations in temperature and composition, especially across their terminators. Their transmission spectra, therefore, carry valuable 3D information.

For example, \cite{Wardenier2024PASP..136h4403W} showed that spectral lines originating from the dayside exhibit increasing blueshifts, caused by the combined effects of spatial distribution and planetary rotation. In contrast, species tracing the nightside, such as \ch{H2O} and TiO, are influenced in the opposite direction. Depending on the planet’s 3D wind structure, this can lead to weaker blueshifts or even redshifts as a function of orbital phase.

%

\bibliographystyle{spphys_ed}
\bibliography{references}

\end{document}